\documentclass[letterpaper, 10 pt, conference]{ieeeconf}

\IEEEoverridecommandlockouts                              

\usepackage{todonotes}
\usepackage[sharp]{easylist}
\usepackage{paralist}
\usepackage{multirow}
\usepackage{tabularx}
\usepackage{url}

\title{\LARGE \bf
Understanding how Software Can Support the Needs of Family Caregivers for Patients with Severe Conditions}

\author{
Angela di Fiore$^{1}$, Francesco Ceschel$^{1}$, Francesca Fiore$^{1}$, Marcos Baez$^{1}$, Fabio Casati$^{1}$ and Giampaolo Armellin$^{2}$
\thanks{$^{1}$Dept of Information Engineering and Computer Science,
        University of Trento, Italy
        {\tt\small first.last at unitn.it}}%
\thanks{$^{2}$CBA Group, Italy
        {\tt\small Giampaolo.Armellin at cba.it}}%
}

\begin{document}

\maketitle
\thispagestyle{empty}
\pagestyle{empty}

\begin{abstract}
In this paper, we report an extensive analysis that we performed in two scenarios where the care relation between doctor and patients are mediated by the relatives of the patients: Pediatric Palliative Care (PPC) and Nursing Homes (NH). When the patients are children or very old adults in the end of life, the provision of care often involve a family caregiver as the main point of contact for the health service. PPC and NH are characterized by emotional complexity, since incurable diseases expose the family caregivers to heavy careload and human distress.
In this paper, we discuss our findings with a novel perspective, focusing on: information, coordination and social challenges that arise by dealing with such contexts; the existing technology as it is appropriated today to cope with them; and what we, as software researchers, can do to develop the right solutions.\footnote{
\copyright2016 IEEE. 
To be published in The 39th International Conference on Software Engineering (ICSE 2017). Personal use of this material is permitted. Permission from IEEE must be obtained for all other uses, in any current or future media, including reprinting/republishing this material for advertising or promotional purposes, creating new collective works, for resale or redistribution to servers or lists, or reuse of any copyrighted component of this work in other works must be obtained from the IEEE. For more details, see the IEEE Copyright Policy.
}  

\end{abstract}

\section{BACKGROUND AND OBJECTIVES}

Pediatric Palliative Care (the end-of-life way of care for children with incurable diseases \cite{miller2015}) and care for elderly people in Nursing Homes are two areas of healthcare characterized by complex social and emotional challenges, in addition to medical ones \cite{telliouglu2014,wiegand2013}.  
Although the patients and diseases are very different, the two scenarios present important similarities:

\begin{enumerate}

\item Patients are typically affected by a chronic condition. This is always the case in Pediatric Palliative Care (PPC), but also Nursing Homes (NHs), due to continuous budget cuts, have been focusing more on care for persons affected by severe conditions (this is the case for Italy, where we performed our studies). Cases of people leaving a nursing home because their condition improved are a minority. For this reason, in both contexts, the treatments mainly focus on maintaining quality of life.

\item These care scenarios are characterized by a \textit{mediated relation} between care professionals and patients where not the patient, but the family caregiver (typically the parent in PPC and the child in NH) is the person that interacts with the care structures and takes decisions. This means that the healthcare institutions take charge of both the patients and their families.

\item Patients are restricted to live permanently in the same building until the end-of-life (this is obvious for NH but often the case also for PPC due to the illness).

\end{enumerate}

An important difference is that in PPC the family also administers the care while in NH the patient is in charge of the NH staff and the family caregiver is mainly for support.
In both scenarios, adults find themselves thrown into uncharted territory, managing a situation that they have never experienced before. 
To make things more emotionally challenging, the transition is often sudden (also in nursing homes, where many admissions come from hospitals), and may provoke tensions within the family, marks the start of a progressive health deterioration\footnote{On a personal note by the authors, this area is very emotionally draining for researchers as well \cite{difiore2016}}.

The relevant literature in this broad area comes from different disciplines. Healthcare studies show that patients with severe conditions are looked after by two typologies of caregivers: formal (health professional) and informal (family) caregivers \cite{weinberg2007}. They are co-producers of care, and their collaboration and mutual trust are essential in the care of the patients \cite{gittell2008}.
However, several studies highlight gaps in communication between formal and informal caregivers, revealing that often the family members have confusion and unanswered questions about the life expectancy of their relatives \cite{schoen2005,kripalani2007}.

Healtcare models such as \textit{continuity of care} focus on integration between caregivers to provide a coherent, transparent and predictable care service \cite{bodenheimer2008}. They support the contribution of all caregivers engaged in the care, by enhancing coordination, and focusing on the needs of the patients and their family \cite{wagner2000}. 
This model stresses the need to work on technologies to facilitate the dynamics among all caregivers for \textit{information continuity} (the need of proper and coherent information), \textit{management continuity} (the need for clear protocols) and \textit{relational continuity} (the need of safe relations and human support) \cite{haggerty2003}.

Most of the existing technology studies (\cite{jeong2008,milligan2012,bossen2013,ruan2010}) focus on  solutions that foster coordination and information exchange issues. 
However, there is an emerging demand for technologies that help informal caregivers in both care and emotional concerns. 
Indeed, families caregivers are especially affected by above average burnout, depression, and stress  \cite{savage2004}
The recognized lack in suitable technological solutions for supporting informal caregivers is a call for actions for software researchers \cite{telliouglu2014}.

In this paper we describe the results of analyses performed over the past two years to understand which role can software applications play in helping people cope with the challenges that these contexts present. We aim in particular at understanding  i) 
which technologies are used today by the caregivers, why, and how effective they are, and 
ii) how can - existing or novel - software applications better address their needs. 
As we will see there is space both for novel use of existing applications as well as new applications, whose requirements were not obvious to us in the beginning (and we try to focus more on these aspects).
We start by describing our analysis method and then report on our findings and recommendations.

\section{METHOD}
To understand the needs we carried out several studies in two different contexts in northern Italy. 
We based our studies mainly on qualitative methodologies, although in the NH case we also developed a data warehouse to analyze populations and processes to the extent allowed by taking information from healthcare IT system, which are very detailed in NHs. 
In PPC, where patients are at home as long as possible, we studied the dynamics between formal and informal caregivers of a PPC network \cite{bossen2013}. We interviewed 15 families, and performed observations in the houses of three families. Data have been collected from July 2015 to March 2016 (by only one researcher, due to the sensitivity of the context). 
A second set of studies focused on six NHs to understand the issues and needs related to family caregivers. 
NHs have a larger population and we had access to a large number of subjects. The visits were conducted in the fall of 2015 and in the spring/summer of 2016, and all attended by at least three researchers, to collect different perspectives and reduce the chances of biases \cite{taylor2016}. 

Specifically, we adopted the following research methods: 
\begin{inparaenum}[\em i)]
\item we carried \textit{in situ} observations in all the contexts, to grasp the organizational and social dynamics that occur among and between family caregivers and care professionals, as well as the communication practices that take places among all the subjects involved, by also creating moments of informal discussion on the emerging issues with our informants \cite{lassiter2005};
\item we interviewed the caregivers - formal and informal - to focus on their emotional discomfort \cite{lamendola2007}, and on the - technological - solutions they adopt to cope with their tasks and communication needs;
\item we involved several formal caregivers in some focus groups to have a deeper understanding of their perspective.
\end{inparaenum}

\newcolumntype{C}[1]{>{\hsize=#1\hsize}X}%

\begin{table*}[t]
\caption{Summary of problems, current practices and opportunities for the technology}
\label{table:summary}
\begin{center}
\begin{tabularx}{\textwidth}{|X|X|}
\hline
\bfseries Contextual problems  & \bfseries Opportunities for technology development\\
\hline
	\multicolumn{2}{|X|}{\bfseries Communications with the care professionals} \\
\hline

\begin{itemize}
\item Lack of transparency and traceability 
\item Lack of clear and (timely) available information 
\item No record of interactions 
\item Overload of the communication channels.
\item Formal and informal exchanges going through the same channel.
\item Lack of mutual trust.
\end{itemize}

\textbf{Tech practices and limitations.}
\begin{itemize}
\item Care activities are scheduled and registered in EHR systems. Information collected is mostly focused on health-related data.
\item Communications are done face to face, via phone, WhatsApp or email (formal / informal with no trace and manually).
\item Facebook pages are used for events and general announcements.  
\end{itemize}

& 

\begin{itemize}
\item  Integration of informal channels with EHR, to keep track of interactions and activities while making use of existing familiar channels.
\item  (Semi-)Automation of the information flow through the different channels - to the extent allowed by the local regulation - to reduce communication overload on the Staff / family.
\item  Expand data collection to aspects of social and psychological well-being, and so accounting for this recurrent information need.
\item  Personalisation of information delivery to key indicators of the patient and preference of the final receiver.
\item  Translation of the information to a format that is understandable in terms of its meaning, implications and curse of action. 
\item  Structured interactions to account for type, priority, sensitivity of information and  so facilitate retrieval and processing. 
\end{itemize}\\

\hline
	\multicolumn{2}{|X|}{\bfseries Communications within the family} \\
\hline

\begin{itemize}
\item Internal coordination issues, and different workload
\item Information not uniformally spread 
\end{itemize}

\textbf{Tech practices and limitations.}

\begin{itemize}
\item Face to face coordination, no trace of performed activities and effort.
\item Sharing via WhatsApp and physical document by one person
\end{itemize}



& 
\begin{itemize}

\item Traceability and visibility of family efforts.
\item Coordination tools that account for the care schedule, and activities of individuals and family as a whole. 
\item Sharing tools that facilitate information flow among family members while still in control of the main responsible. 

\end{itemize}

\\
\hline
	\multicolumn{2}{|X|}{\bfseries Social support for the family} \\
\hline
\begin{itemize}
\item Social isolation
\item Emotional distress
\item Need of feeling understood
\end{itemize}

\textbf{Tech practices and limitations.}

\begin{itemize}
\item Social support groups enabled via Whatsapp and Facebook private groups  but  problems finding  relevant groups /peers.
\item Psychological consultations, though not available in all institutions.
\end{itemize}



& 
\begin{itemize}
\item Widening the support network, facilitating the discoverability of relevant support groups.
\item Organizing online peer support networks with existing technology, (possibly) moderated by an expert.
\item (Self-)Coaching systems implementing existing successful programs to improve the psychological, emotional and social well-being. 
\item Monitoring of the psychological well-being of the relatives
\end{itemize}

\\
\hline
	\multicolumn{2}{|X|}{\bfseries Education} \\
\hline
\begin{itemize} 
\item Confusion in what to do and expect.
\item Lack of medical/care knowledge and 
medical language.
\end{itemize}
\textbf{Tech practices and limitations.}
\begin{itemize}
\item ``Doctor google" and facebook groups leading to inconsistent info.
\item Exchanges with other caregivers, face-to-face though not optimal. 
\end{itemize}



& 
\begin{itemize}
\item Peer-to-peer networks that allow sharing of practices and experiences, (possibly) supported by the moderation of medical experts. 
\item Facilitating access to portals with certified information.
\item Expert support systems to help family in care activities.
\end{itemize}
\\
\hline

\end{tabularx}
\end{center}
\end{table*}

\section{FINDINGS}
The analysis of the gathered data show that there are four main areas of problems where technology can be of help (See Table \ref{table:summary}).

\noindent \textit{1. Communication with the care professionals}: this emerged as a major issue in both PPC and NH. In PPC, formal and informal interactions (e.g., cute photos of  and information on treatment) travel on the same channel, which is typically Whatsapp.
Whatsapp enhances collaboration between formal and informal caregivers, allowing real-time exchange of clinical documents (such as discharge letters and tests results) and quick remote medical consultations. Usually, the mother sends a photo or a video that shows the exacerbation of a medical condition to the members of the PPC unit by asking what to do.

While this has many positive aspects (chat software is free, easy, fits into the natural daily behavior and everybody uses it), it also creates a problem in terms of lack of traceability and monitoring, unclear management of privacy, as well as communication overload (chats happen frequently and at any time) which results in the risk of losing important messages. 
In NH the interaction is by phone or F2F. The same problem of overload exists here, but in NH they complement much bigger problems which are i) lack of trust in the abilities and willingness of NH staff to provide care, and ii) belief that the loved one may be mistreated, due to news of criminal behaviors in NH that is sometimes reported in the national news. Furthermore, the family also feels a lack of clear and timely information.

The interesting, and for us unexpected aspect in NH is that the staff, due to the interaction overload and frustrating feeling of lack of trust, are extremely supportive of any system that provides transparency into the life in a NH. Notice that, while the interaction problems with a given family tend to reduce over time, most NH (as we understood from the warehouse data) have a turnover ranging from 20 to 40\% per year. This means that there are always new families to cope with. Furthermore, we learned that the staff interacts differently with the families based on their classification of "personas": with some family member they are more open and direct, with others there are more careful in the information they reveal, because of the perceived risk of over-reactions. 
Finally, an important finding was that NHs already have an information system which they populate in great detail, every day for every resident. So most of the information needed to provide information and transparency is there, though not in the form that can be understood by relatives (and it may not always be wise to reveal them automatically). 

\noindent \textit{2. Interaction within families}: PPC and NH both create very strong tensions within the family, mostly related to different emotional reactions to the problem or to disagreement in how to handle it. For example, in NH the children of the resident sometimes disagree on the choice of taking the parent to a NH, on who should go visit and on who foots the bills. We also observed frustration by family caregivers who visit more often towards those who come less often.
The technology used to involve the family more in this case is again chat software, used to both inform the whole family on the situation but in part also as a tool to make relatives feel a bit "guilty" because they are not visiting as much.

\noindent \textit{3. Social support for the Family}. 
The transition to care for a relative in chronic condition is always very painful. 
In addition, this transition often brings with it a social isolation because of the need (or desire) to spend time with the loved ones, but also because it can become difficult to spend time with people that do not understand what you are going through.

Social support is known as a useful method for coping with traumatic situations. In PPC, family caregivers rely on Facebook groups to connect with other parents who experience the same situation from allover the world, allowing for peer-to-peer conversations to find social support, and to receive useful suggestions.
However, the specificity of each illness (which in many cases is some form of rare disease) makes it difficult to find people who are living an experience similar to yours.
In NH the problems are more "standardized" but the family caregivers are often relatively old themselves and do not use technology beyond, sometimes, email and chat.\\
\noindent \textit{4. Education and Managing Expectations}. 
A huge source of problems and misunderstanding between family and professionals is the lack of knowledge and wrong assumptions on i) how the patient's health will evolve and ii) what the healthcare system can do about it. Very often family believes the action of the professionals should be care or rehabilitation, but this is often impossible due to the medical conditions of the person or, in the NH case, to lack of staff for performing, for example, what would be a complex physical rehab program. 
The problem of erroneous expectations is manifested by the fact that often the patient is not aware that their situation is permanent, even in the NH case.
In this case the technology used today is essentially web browsing and searching for information, but this is sometimes the cause of the problem which is indeed fostered by the use of diverse and inconsistent sources on the Internet. For example, in NH, because there are so many "types" of NH in different countries with very different population, one may find information on the internet that does not apply to the NH at hand, but mistakenly believe it does. The same is true for many aspects of care (such as prescriptions of medicine).

\section{OPPORTUNITIES FOR TECHNOLOGY DEVELOPMENT AND ADOPTION}

In this section we summarize opportunities for novel technology (or usages of existing technology that fit the problems at hand) for each problem category. We focus on what we found more interesting and surprising and omit discussions on security, privacy, data integration, usability, and other concerns the reader may expect.

In family-staff interactions, by looking at the NH scenario it becomes apparent that a portal that allows relatives to view the status and activities of the relatives is both feasible and useful.
It is feasible because NH staff already fills detailed information on the residents in an IT system, for internal reasons. This means that much of the information is already there. It is feasible also because the staff does want more transparency.
And it is useful as relatives requested such information (and indeed they do so today, by phone).

Three key requirements emerged from the analysis:
The staff segments the relatives into “personas” that react to news in different ways and with whom today they use different communication strategies, and so the software must support this.
Information also needs to be classified according to the level of approval required before sending it to the relatives: some information can be sent to all relatives automatically (e.g., the menu of the day, the wake up time, etc), some information requires explicit prior approval that it is ok to send, and other information needs to be edited/rewritten to avoid unnecessary concerns (The latter case also depends on the personas, and it may be different for new or “experienced” relatives). 
Because the relative might ask for clarifications, it is important that each staff member can have easy access to exactly what the relative has seen in the portal.

An additional observation that emerged is that NHs today do not really collect information about subjective wellbeing (of residents and relatives) while it would the important to do so given that quality of life is a key aspect of care. 

In PPC, the opportunity lies more in taking the instant messaging paradigm and (semi-automatically) extracting  messages related to coordination and administration of care. 
Ad hoc applications and a portal like in NH may also be proposed but it is unclear that they would be adopted, because the PPC care network is wide and ad hoc applications become effective if everybody uses them.

For interactions within the family, an opportunity that emerged is the obvious extension of the portal above, where the entire family can be given access to. But what appeared even more strongly is the need to involve the family members beyond the family caregiver using the instruments they already use. For example, grandchildren of residents can be involved by pushing “involving” images or information to chat (as we experimented with telegram bots for telegram users) or Instagram, as well as add events and visit schedule to a calendar.
In those PPC networks where a dedicated app is not be adopted for the reason stated above, a way to easily map whatsapp exchanges into calendars would already be beneficial. 

Opportunities for social support and education are instead more in terms of reusing existing technology but with better aggregation of content and people. For example, PPC would benefit for a single place that contains a set of forums, one for each rare disease, so that parents know where to go. Similarly for NH adults would benefits from illness-specific forums as well as forums related to NHs in their region, both for support but also to compare care practices and manage their expectations. All this can be integrated into a same portal and app, though the challenges here are in terms of content organization.

Table \ref{table:summary} summarizes the common points for each scenario. In summary, there are several directions in which we as software researchers and engineers can contribute to make a difference in this difficult and stressful context, essentially by enabling easy access to personalized information that provide transparency into care processes and information relevant to the physical and care conditions of our loved one.

\textbf{\textit{Verifiability}:}
Our studies are based on a total of 35 interviews, 40 days of observation, 4 focus groups, a warehouse with data on over 4000 subjects, and document analysis of processes and health records. The work has received three ethical approvals (as available at: http://bit.ly/2eRFte8). Our data collection can be verified through a formal consultation (since data are sensitive, the consultation must be allowed by a formal permission of our ethical committees).

\section*{Acknowledgment}
This work is supported by the Trentino project "Collegamenti" that is being funded by the Province of Trento (l.p. n.6-December 13rd 1999).
This project has also received funding from the European Union's Horizon
2020 research and innovation programme under the Marie
Skłodowska-Curie grant agreement No 690962.


\addtolength{\textheight}{-12cm}   
                              



\bibliographystyle{IEEEtran}
\bibliography{tnhappiness,tnreferences,icse2017}

\end{document}